\documentclass{aa}
\usepackage{graphicx}
\begin{document}
\title{A study of the two northern open clusters \\
       \hspace{1cm} NGC~1582 and NGC~1663
       \thanks{Based on observations carried out at Mt Ekar, Asiago, Italy},
       \thanks{Data is only available in electronic form at the CDS via 
               anonymous ftp to {\tt cdsarc.u-strasbg.fr (130.79.128.5)} or 
	       via {\tt http://cdsweb.u-strasbg.fr/cgi-bin/qcat?J/A+A//}
	       }
       }
       \author{{}Gustavo Baume\inst{1,2},
		 Sandro Villanova\inst{1}
		 \and
		 Giovanni Carraro\inst{1}
	      }

   \offprints{baume@pd.astro.it}

   \institute{Dipartimento di Astronomia, Universit\`a di Padova,
              Vicolo Osservatorio 2, I-35122 Padova, Italy
         \and
             Facultad de Ciencias Astron\'omicas y Geof\'{\i}sicas de la
	     UNLP, IALP-CONICET, Paseo del Bosque s/n, La Plata, Argentina\\
	     }

   \date{Received **; accepted **}

\abstract{
We present CCD $UBV(I)_C$ observations obtained in the field of the previously 
unstudied northern open clusters NGC~1582 and NGC~1663. For the former, we also
provide high-resolution spectra of the brightest stars and  complement our data 
with Two-Micron All-Sky-Survey (2MASS) near-infrared photometry and with 
astrometric data from the Tycho-2 catalog.\\ 
From the analysis of all these data, we argue that NGC~1582 is a very poor, 
quite large and heavily contaminated open cluster. It turns out to have a 
reddening $E_{B-V} = 0.35 \pm 0.03$, to be situated $1100 \pm 100$~pc from the 
Sun and to have an age of $300 \pm 100$~Myr. \\
On the other hand, we were not able to unambiguously clarify the nature of 
NGC~1663. By assuming it is a real cluster and from the analysis of its
photometric diagrams, we found a color excess value $E_{B-V} = 0.20$, 
an intermediate age value ($\sim 2000$~Myr) and a distance of about $700$~pc. 
The distribution of the stars in the region however suggests we are probably 
facing an open cluster remnant. As an additional result, we obtained aperture 
photometry of three previously unclassified galaxies placed in the field of 
NGC~1663 and performed a preliminary morphological classification of them.
}

\authorrunning{Baume et al.}
\titlerunning{NGC 1582 and NGC 1663}

\maketitle
%

\section{Introduction}

$\hspace{0.5cm}$
This study is part of a long term project aimed at providing accurate CCD 
photometry for poorly known or unstudied northern open clusters (Carraro 2002a, 
and references therein). Here we focus our attention on the open clusters 
NGC~1582 and NGC~1663. Both objects have never been studied before, apart from
the identification and a preliminary estimate of their angular size. Their basic 
data are summarized in Table~1. \\

\noindent
By inspecting the finding charts of these two clusters (see Figs. 2 and 10), we 
notice mainly that:
\begin{itemize}
\item NGC~1582 lies in a region where several bright stars are present, many of 
      them with a Henry Draper (HD) classification. The cluster appears as a 
      weak concentration of a small group of bright stars well mixed with the 
      very rich Galactic disk field star population toward its direction. This 
      renders it difficult to study objects like this, and this is -we guess- 
      the main reason for which this object has been almost neglected up to now; 
\item NGC~1663, which is located fairly high above the galactic plane, suffers 
      from less contamination, and resembles the kind of objects recently 
      suggested by Bica et al. (2001) to be Probable Open Cluster Remnants 
      (POCRs).
\end{itemize}

\noindent
In this study we would like to address the issue of the real nature of these two 
objects and to provide the first estimate of their fundamental parameters, 
namely distance, reddening, size and age. Therefore, we performed multicolor 
$UBV(I)_C$ photometry for both, and high-resolution spectroscopy for some stars in the NGC 
1582 field. We also complement and cross-correlate our data with the 2MASS 
catalog and with proper motions from the Tycho-2 catalog (H{\o}g et al. 
2000), whenever available.\\

\noindent
The plan of this study is as follows: In Sect.~2 we briefly present the 
observations and data reduction. In Sect.~3 and Sect.~4 we illustrate our 
analysis and results for NGC 1582 and NGC 1663, respectively. Sect.~5 is 
dedicated to a preliminary analysis of three previously unclassified galaxies in 
the field of NGC~1663. Finally, in Sect.~6 we draw our conclusions. 
\\

\begin{table}
\caption{{}Basic data of the observed objects.}
\fontsize{8} {10pt}\selectfont
\begin{tabular}{ccccc}
\hline
\multicolumn{1}{c} {$Name$} &
\multicolumn{1}{c} {$\alpha_{2000}$}  &
\multicolumn{1}{c} {$\delta_{2000}$}  &
\multicolumn{1}{c} {$l$} &
\multicolumn{1}{c} {$b$} \\
\hline
NGC 1582 & 04:32:15.4 & +43:50:43 & $159.30^{\circ}$ & $ -2.89^{\circ}$ \\
NGC 1663 & 04:49:24.3 & +13:08:27 & $185.92^{\circ}$ & $-19.65^{\circ}$ \\     
\hline
\end{tabular}
\end{table}

\section{Observations and data reduction}

\subsection{Photometry}

$\hspace{0.5cm}$
CCD $UBV(I)_C$ observations were carried out with the AFOSC camera at the 
1.82 m Copernico telescope of Cima Ekar (Asiago, Italy), in the photometric 
night of November 8, 2002. AFOSC samples a $8^\prime.14\times8^\prime.14$ field 
in a $1K\times 1K$ nitrogen-cooled thinned CCD. 

\begin{table} 
\fontsize{8} {10pt}\selectfont
\caption{Journal of observations of NGC 1582, NGC 1663 and standard star fields 
together with calibration coefficients (November 8, 2002).} 
\begin{tabular}{ccccccc} 
\hline 
\multicolumn{1}{c}{Field}         & 
\multicolumn{1}{c}{Filter}        & 
\multicolumn{3}{c}{Exposure time} & 
\multicolumn{1}{c}{Seeing}        &
\multicolumn{1}{c}{Airmass}       \\
 & & \multicolumn{3}{c}{[sec.]} & [$\prime\prime$] & \\ 
\hline 
 NGC 1582       & U &  900x2 & 180 & 20 & 2.5 & 1.196 \\ 
                & B &  600   &  60 & 10 & 2.3 & 1.230 \\ 
                & V &  300   &  30 &  5 & 2.4 & 1.255 \\ 
                & I &  300   &  30 &  5 & 2.1 & 1.274 \\
\hline
 NGC 1663       & U &  900x2 &   - &  - & 2.5 & 1.348 \\ 
                & B &  600   &  60 &  - & 2.4 & 1.215 \\ 
                & V &  300   &  30 &  3 & 2.2 & 1.188 \\ 
                & I &  300   &  30 &  3 & 2.0 & 1.190 \\
\hline
PG 0231+051     & U &  800   &     &    & 2.5 & 1.348 \\
                & B &  300   &     &    & 2.4 & 1.324 \\ 
                & V &   60   &     &    & 2.2 & 1.316 \\ 
                & I &   90   &     &    & 2.2 & 1.315 \\ 
\hline
PG 2213-006     & U &  600   &     &    & 2.5 & 1.447 \\
                & B &  150   &     &    & 2.3 & 1.457 \\ 
                & V &   30   &     &    & 2.3 & 1.465 \\ 
                & I &   30   &     &    & 2.3 & 1.472 \\ 
\hline 
\hline
Calibration     & \multicolumn {3}{l}{$u_1 = +3.861 \pm 0.015$}     & \multicolumn {3}{l}{$b_1 = +1.602 \pm 0.004$} \\
coefficients    & \multicolumn {3}{l}{$u_2 = -0.142 \pm 0.022$}     & \multicolumn {3}{l}{$b_2 = +0.038 \pm 0.006$} \\
                & \multicolumn {3}{l}{$u_3 = +0.58$}                & \multicolumn {3}{l}{$b_3 = +0.29$}            \\
		& \multicolumn {3}{l}{$v_{1bv} = +1.003 \pm 0.014$} & \multicolumn {3}{l}{$i_1 = +1.691 \pm 0.044$} \\
		& \multicolumn {3}{l}{$v_{2bv} = -0.016 \pm 0.018$} & \multicolumn {3}{l}{$i_2 = +0.057 \pm 0.043$} \\
		& \multicolumn {3}{l}{$v_3 = +0.16$}                & \multicolumn {3}{l}{$i_3 = +0.08$}            \\
		& \multicolumn {3}{l}{$v_{1vi} = +1.002 \pm 0.016$} & \\
		& \multicolumn {3}{l}{$v_{2vi} = -0.013 \pm 0.016$} & \\
\hline
\end{tabular}
\end{table}

\begin{figure}
\centering
\includegraphics[width=8cm,height=11cm]{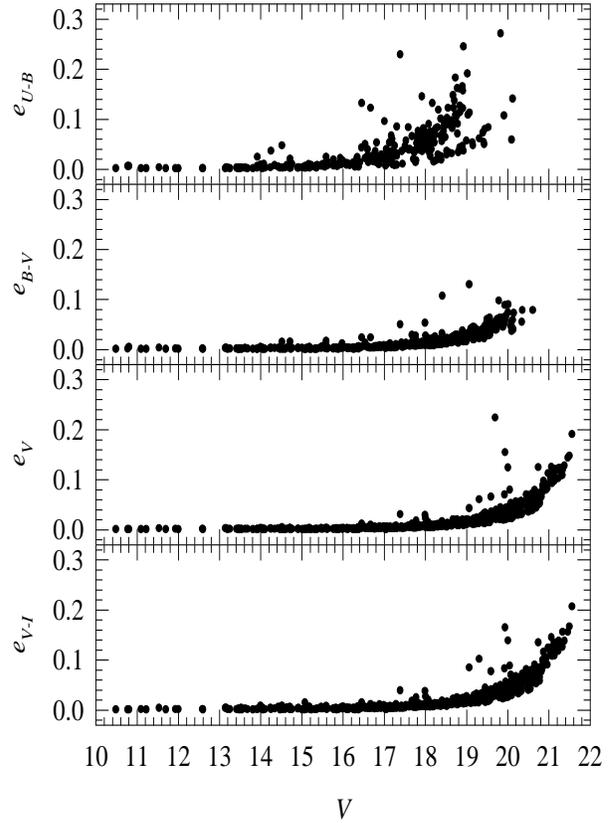}
\caption{{} {Photometric errors in $V$ magnitude and $U-B$, $B-V$ and $V-I$ 
colours as a function of $V$.}}
\end{figure}

\begin{figure*}
\centering
\includegraphics[width=16cm]{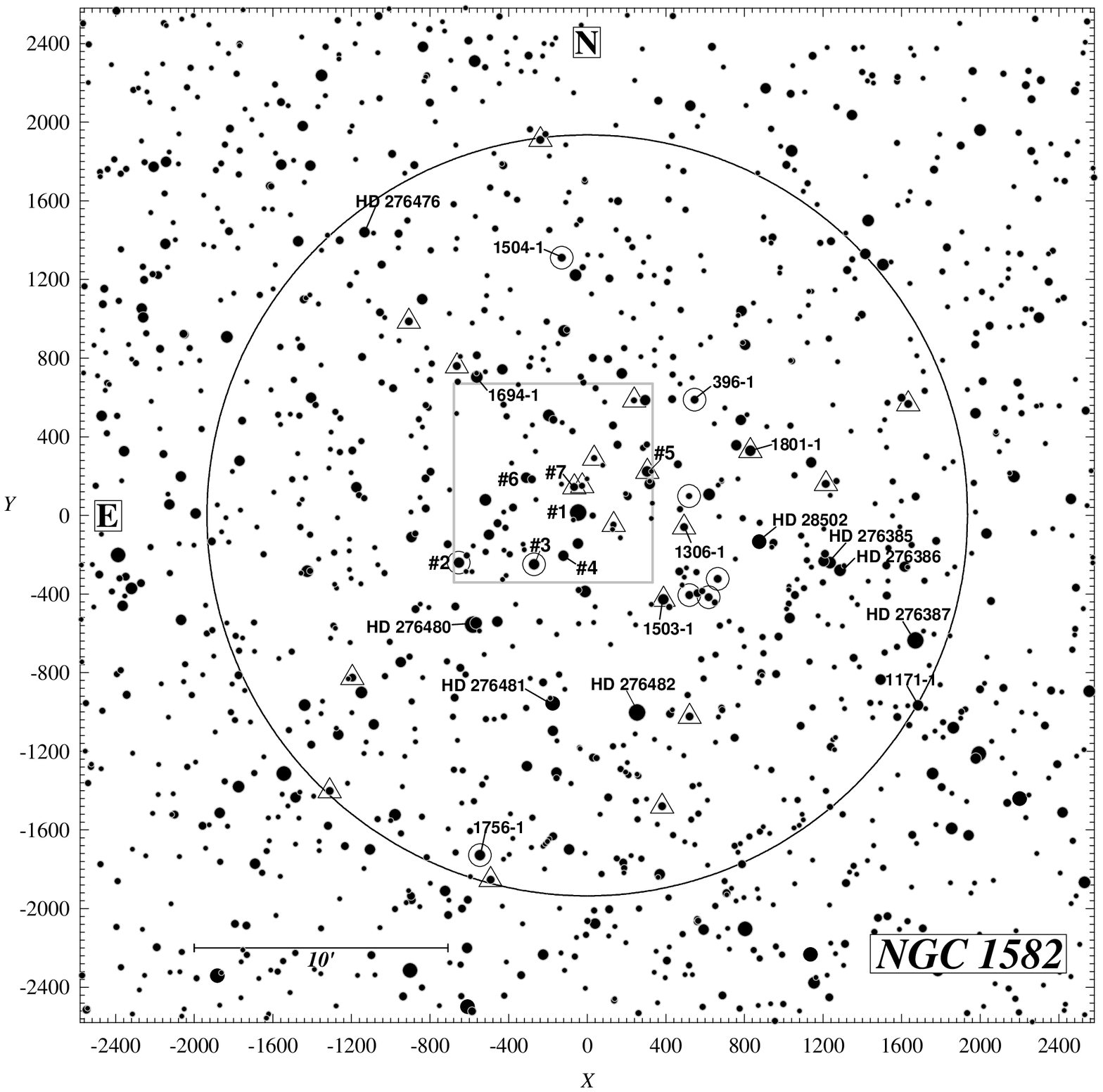}
\caption{{} Finding chart of the NGC~1582 region ($40^{\prime} \times 
40^{\prime}$ and $J$ filter). The black solid circle, $15^{\prime}$ in radius, 
indicates the adopted angular size for the cluster (see Sect. 3.1 and Fig. 3). 
The grey square indicates the area covered from Asiago. Adopted cluster members 
and probable cluster members are shown with small circles and triangles, 
respectively. For a coordinate reference, the center ($X = 0$; $Y=0$) 
corresponds to the cluster coordinates (see Table~1) and each $X$-$Y$ unit is 
$0^{\prime\prime}.465$.}
\end{figure*}

\renewcommand{\thefootnote}{\dag}
\noindent
Details of the observations are listed in Table~2, where the observed fields are 
reported together with the exposure times, the typical seeing values and the 
air masses. Figs. 2 and 10 show the finding charts of the NGC 1582 and NGC 1663
regions respectively, indicating the covered areas and the object angular sizes.
The data has been reduced with the 
IRAF\footnote{IRAF is distributed by NOAO, which are operated by AURA under 
cooperative agreement with the NSF.} 
packages CCDRED, DAOPHOT, and PHOTCAL using the point spread function (PSF)
method (Stetson 1987). The calibration equations obtained by observing Landolt 
(1992) PG 0231+051 and PG 2213-006 fields during the night, are: 

\begin{center}
\begin{tabular}{lc}
$u = U + u_1 + u_2 (U-B) + u_3 X$         & (1) \\
$b = B + b_1 + b_2 (B-V) + b_3 X$         & (2) \\  
$v = V + v_{1bv} + v_{2bv} (B-V) + v_3 X$ & (3) \\  
$v = V + v_{1vi} + v_{2vi} (V-I) + v_3 X$ & (4) \\  
$i = I + i_1 + i_2 (V-I) + i_3 X$         & (5) \\
\end{tabular}
\end{center}

\noindent
where $UBVI$ are standard magnitudes, $ubvi$ are the instrumental ones, $X$ is 
the airmass and the used coefficients are presented at the bottom of Table~2. 
As for $V$ magnitudes, when $B$ magnitude was available, we used 
expression (3) to compute them, elsewhere expression (4) was used. The standard 
stars in these fields provide a very good color coverage, essential to obtain 
reliable transformations. For the extinction coefficients, we assumed the 
typical values for the Asiago Observatory ($u_3$, $b_3$, $v_3$ and $i_3$ values
in Table~2, 
Desidera et al. 2002\footnote{http://www.pd.astro.it/Asiago/2000/2300/2310.html}). 
Photometric global errors have been estimated following Patat \& Carraro (2001) 
and their trends against $V$ magnitude are shown in Fig.~1. \\

\subsection{Spectroscopy}

$\hspace{0.5cm}$
High-resolution ($R \approx 30,000$) 
spectra of 10 stars (see Table~3 and Fig.~5) in the field of NGC 1582 were obtained during 
the nights of January 14-15 and February 14, 2003, using the REOSC Echelle 
Spectrograph attached to the 1.82~m telescope of Asiago Astronomical 
Observatory. This instrument works with a Thomson 1024$\times$1024 CCD and the 
allowed wavelength coverage is approximately $4500 - 6650$ \AA. Details on this 
instrument are given in Munari \& Zwitter (1994) and on the Asiago
Obervatory Home page\footnote{http://www.pd.astro.it/Asiago/2000/2300/2320.html}. \\
The exposure times were 45 minutes for all the stars. In order to improve the 
signal-to-noise ratio, two exposures were taken for each star reaching at the 
end $S/N$ values up to 70. The data have been reduced with the IRAF package 
ECHELLE using thorium lamp spectra for wavelength calibration purposes. By 
comparing final known sky line positions along the spectra we derived an error 
of about 0.01 \AA.\\

\begin{figure*}
\centering
\includegraphics[width=15cm]{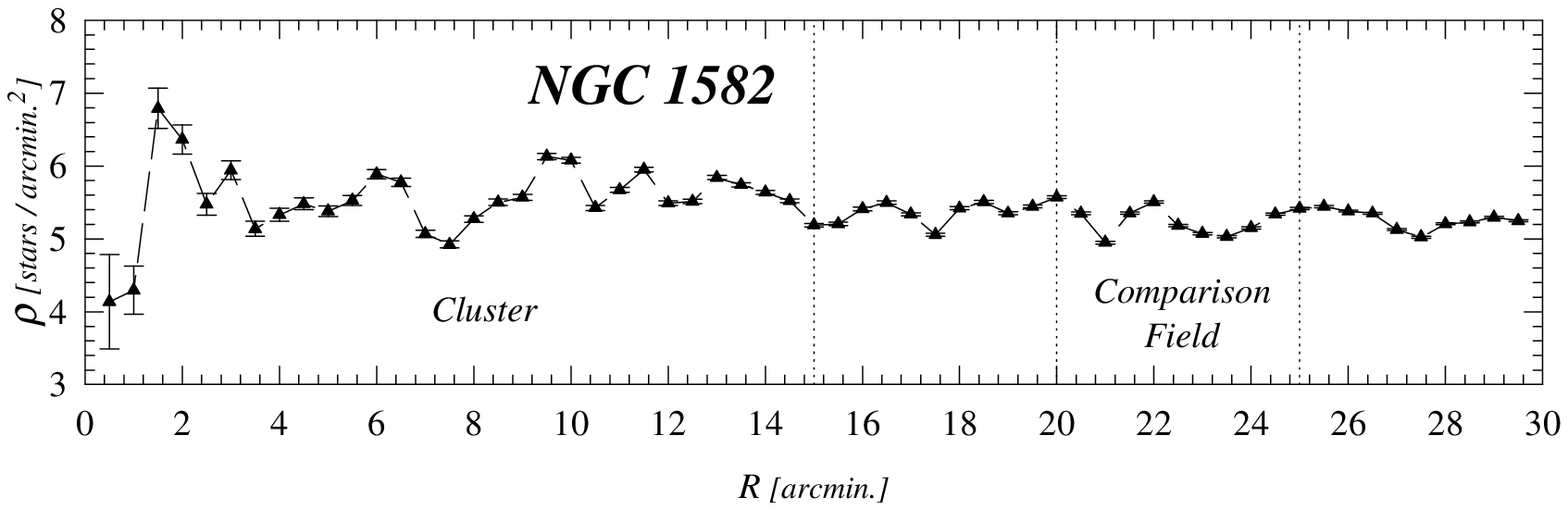}
\caption{{}Stellar surface density profile in the region of NGC~1582 as a function of the 
radius. Data have been taken from 2MASS catalog. Dotted lines indicate the 
adopted limits for the cluster and for the comparison field related with 
Fig.~8b.}
\end{figure*}

\begin{figure}
\centering
\includegraphics[width=7cm]{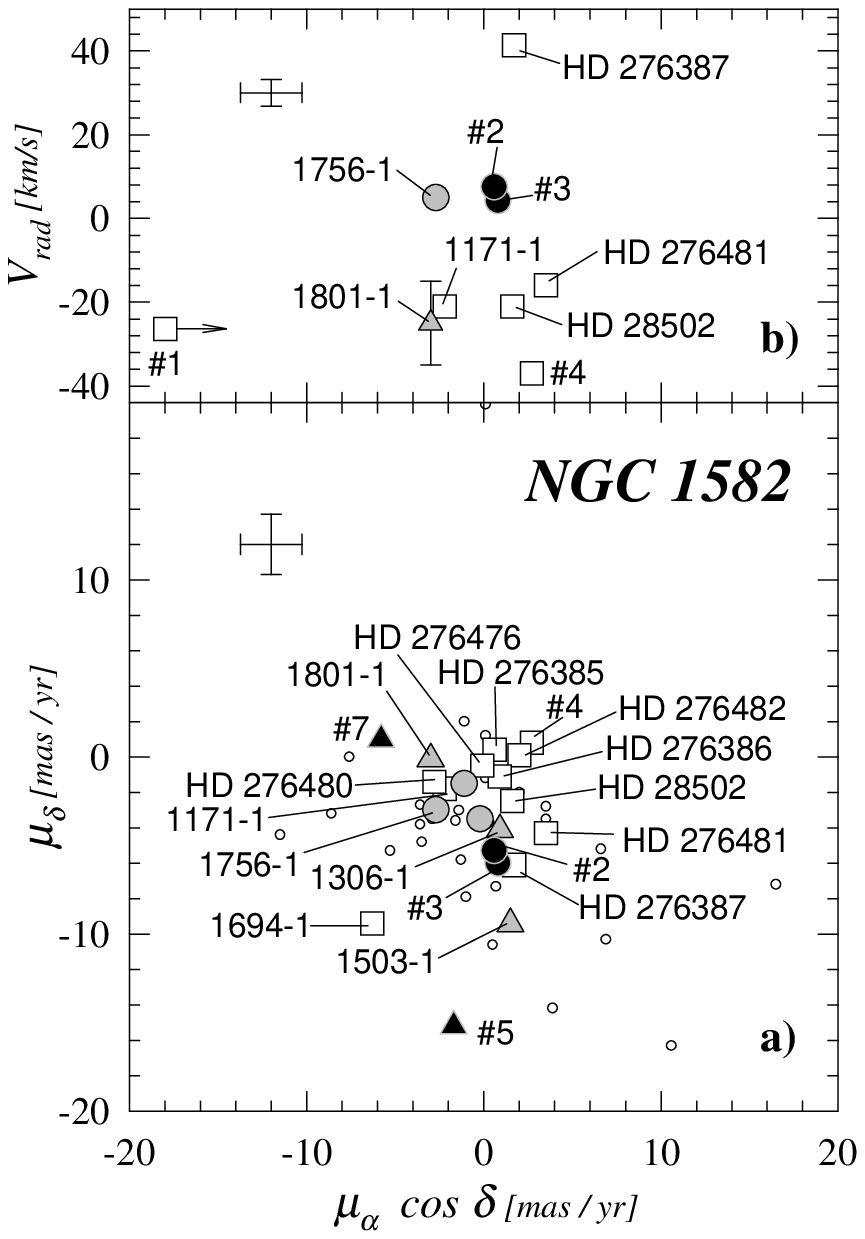}
\caption{{} {\bf a)} Vector point plot for the stars $20^{\prime}$ around the 
center of NGC 1582 (data from HD Extension Charts and Tycho-2 catalogs). 
{\bf b)} Radial velocities obtained from Echelle spectra. See captions of 
Figs.~6 and 8 for the meaning of the symbols. The crosses indicate the mean 
error values.}
\end{figure}

\noindent
Spectra were classified mainly on the basis of the intensity of the Fe, H  and 
He lines -depending on the spectral type- and by comparing them with 
spectrophotometric standards. The results are shown in Fig.~5, where we plot the 
Echelle spectra of the 10 measured stars degraded to a resolution of about 
90~\AA/mm. A few main lines are also shown. These spectra have been then 
compared with stellar libraries from Montes et al. (1997, 1999) and Jaschek \& 
Jaschek (1987) to obtain estimates of spectral types and luminosity classes.\\
 
\noindent
Individual radial velocity values were obtained with the RVIDLINES task using up 
to 10 lines. Typical errors in the radial velocity amount to 3.0~km/s. The 
resulting classification together with the radial velocity values are listed in 
Table~3. \\ 

\section{NGC 1582}

$\hspace{0.5cm}$
NGC~1582 (= OCL 407 = C0428+43) is commonly considered an open cluster with a 
diameter in the range of $15^{\prime}$ (Dias et al. 2002) to $37^{\prime}$ 
(Lyng\aa\ 1987). Our CCD observations only cover the cluster central region 
(see Fig.~2). Therefore, in order to better probe the cluster size and nature, 
we complemented our data with near IR photometry from the 2MASS catalog. \\

\subsection{Stellar counts}

$\hspace{0.5cm}$
\renewcommand{\thefootnote}{\ddag}
To derive the radial stellar surface density we adopted the cluster center 
given by Dias et al. (2002, see Table~1), and we use: {\it a)} the 
corresponding DSS-2\footnote{Second generation Digitized Sky Survey, 
{\tt http://www.eso.org/dss}} 
red image of $40^{\prime} \times 40^{\prime}$, and {\it b)} all the stars from 
the 2MASS catalog inside a circle of radius 30$^{\prime}$, centered on the 
adopted cluster center. The computation is done by performing star counts inside 
increasing concentric rings $1^{\prime}$ wide around the cluster center and then 
dividing by their respective surfaces. The density profiles obtained in the two 
cases are very similar. For the sake of illustration, only the one derived from 
2MASS data is shown in Fig.~3. By inspecting this plot we notice that the 
cluster appears as a weak over-density up to $\sim 15^{\prime}$ from the center, 
where the stellar density reaches the field value. As anticipated, the 
cluster turns out to be much larger than the area we covered with our CCD 
observations and we adopt $15^{\prime}$ as the radius. This way, the obtained 
diameter has a value close to the previous qualitative suggestion by 
Lyng\aa\ (1987). \\

\begin{table*}
\caption{{} Some bright stars in the region of NGC~1582. }
\fontsize{8} {10pt}\selectfont
\begin{center}
\begin{tabular}{rlrcr@{.}lr@{.}lr@{.}lr@{.}lr@{.}lr@{$\pm$}lll}
\hline
 \# & 2MASS ID.       & \multicolumn{1}{c}{$X$} & $\alpha_{2000}$ & \multicolumn{2}{c}{$V$} & \multicolumn{2}{c}{$B-V$} & \multicolumn{2}{c}{{}$E_{B-V}$} & \multicolumn{2}{c}{$J$} & \multicolumn{2}{c}{$J-K$} & \multicolumn{2}{c}{{}$\mu_{\alpha} \cos(\delta)$} & $SC$ & Memb. \\
    & Tycho-2 ID.     & \multicolumn{1}{c}{$Y$} & $\delta_{2000}$ & \multicolumn{2}{c}{}    & \multicolumn{2}{c}{$U-B$} & \multicolumn{2}{c}{{}$E_{U-B}$} & \multicolumn{2}{c}{}    & \multicolumn{2}{c}{}      & \multicolumn{2}{c}{{}$\mu_{\delta}$}              &      &       \\
    & HD/GSC ID.      &                         &                 & \multicolumn{2}{c}{}    & \multicolumn{2}{c}{$V-I$} & \multicolumn{2}{c}{{}$E_{V-I}$} & \multicolumn{2}{c}{}    & \multicolumn{2}{c}{}      & \multicolumn{2}{c}{{}$V_R$}                       &      &       \\
\hline
 -- &    --           & 1671.0 & 04:31:3.24 &  8&61$_T$             &  1&59$_T$ & 0&26 & \multicolumn{2}{c}{--} & \multicolumn{2}{c}{--} &   1.7 & 1.2        & K7 V       & $nm$  \\
    & TYC 2892-803-1  & -636.3 & 43:45:55.0 &  \multicolumn{2}{c}{} &  1&40$_C$ & 0&19 & \multicolumn{2}{c}{}   & \multicolumn{2}{c}{}   &  -6.1 & 1.2        &            &       \\
    & HD 276387       &        &            &  \multicolumn{2}{c}{} &  1&94$_C$ & 0&33 & \multicolumn{2}{c}{}   & \multicolumn{2}{c}{}   &  41.4 & 1.0        &            &       \\
[1 ex]
 -- & J0431376+434944 &  876.0 & 04:31:37.6 &  8&69$_T$             &  0&23$_T$ & 0&33 &  8&06                 &  0&15                 &   1.6 & 1.0          & B8 V p     & $nm$  \\
    & TYC 2892-510-1  & -134.4 & 43:49:44.6 &  \multicolumn{2}{c}{} & -0&06$_C$ & 0&24 & \multicolumn{2}{c}{}  & \multicolumn{2}{c}{}  &  -2.5 & 1.1          &            &       \\
    & HD 28502        &        &            &  \multicolumn{2}{c}{} &  0&30$_C$ & 0&56 & \multicolumn{2}{c}{}  & \multicolumn{2}{c}{}  & -21.0 & 3.0          &            &       \\
[1 ex]
 -- & J0432225+434316 & -175.9 & 04:32:22.5 &  9&33$_T$             &  0&20$_T$ & 0&33 &  8&71                 & -0&02                 &   3.5 & 1.0          & B6 V       & $nm$  \\
    & TYC 2892-1450-1 & -957.8 & 43:43:16.7 & \multicolumn{2}{c}{}  & -0&18$_C$ & 0&25 & \multicolumn{2}{c}{}  & \multicolumn{2}{c}{}  &  -4.3 & 1.0          &            &       \\
    & HD 276481       &        &            & \multicolumn{2}{c}{}  &  0&32$_H$ & 0&46 & \multicolumn{2}{c}{}  & \multicolumn{2}{c}{}  & -16.0 & 3.0          &            &       \\
[1 ex]
  1 & J0432173+435049 &  -45.9 & 04:32:17.3 & 10&77                 &  2&01     & \multicolumn{2}{c}{--} &  6&75                &  1&22                 & \multicolumn{2}{l}{~~~~~~~--} & M3-7 V & $nm$  \\
    & TYC 2892-1209-1 &   14.1 & 43:50:49.2 & \multicolumn{2}{c}{}  &  2&52     & \multicolumn{2}{c}{--} & \multicolumn{2}{c}{} & \multicolumn{2}{c}{}  & \multicolumn{2}{l}{~~~~~~~--} &        &       \\
    & GSC 02892-01209 &        &            & \multicolumn{2}{c}{}  &  2&47     & \multicolumn{2}{c}{--} & \multicolumn{2}{c}{} & \multicolumn{2}{c}{}  & -26.4 & 1.6                   &        &       \\
[1 ex]
  2 & J0432433+434847 & -654.1 & 04:32:43.3 & 10&80                 &  0&33     & 0&35 & 10&02                 &  0&17                 &  0.6 & 1.4           & A0 V *     & $m$ \\
    & TYC 2892-354-1  & -241.6 & 43:48:47.7 & \multicolumn{2}{c}{}  &  0&32     & 0&34 & \multicolumn{2}{c}{}  & \multicolumn{2}{c}{}  & -5.3 & 1.4           &            &       \\
    & HD 276479       &        &            & \multicolumn{2}{c}{}  &  0&46     & 0&47 & \multicolumn{2}{c}{}  & \multicolumn{2}{c}{}  &  7.5 & 2.2           &            &       \\
[1 ex]
 -- & 0432381+433716  & -545.4 & 04:32:38.1 & 10&81$_T$              &  0&22$_T$ & 0&34 & 10&23                &  0&06                 & -2.7 & 1.4           & B7 V       & $m$ \\
    & TYC 2892-1756-1 &-1729.7 & 43:37:16.0 & \multicolumn{2}{c}{}   & -0&12$_C$ & 0&25 & \multicolumn{2}{c}{} & \multicolumn{2}{c}{}  & -3.0 & 1.3           &            &       \\
    & GSC 02892-01756 &        &            & \multicolumn{2}{c}{}   &  0&30$_C$ & 0&42 & \multicolumn{2}{c}{} & \multicolumn{2}{c}{}  &  4.9 & 1.0           &            &       \\
[1 ex]
  3 & J0432269+434845 & -270.2 & 04:32:26.9 & 11&09                 &  0&31     & 0&38 & 10&39                 &  0&19                 &  0.8 & 1.5           & B9 V p     & $m$ \\
    & TYC 2892-1365-1 & -249.8 & 43:48:45.5 & \multicolumn{2}{c}{}  &  0&10     & 0&30 & \multicolumn{2}{c}{}  & \multicolumn{2}{c}{}  & -6.0 & 1.5           &            &       \\
    & GSC 02892-01365 &        &            & \multicolumn{2}{c}{}  &  0&39     & 0&45 & \multicolumn{2}{c}{}  & \multicolumn{2}{c}{}  &  4.2 & 2.1           &            &       \\
[1 ex]
  4 & J0432205+434906 & -121.1 & 04:32:20.5 & 11&22                 &  0&29     & 0&40 & 10&59                 &  0&11                 &   2.7 & 1.5          & B8 V       & $nm$  \\
    & TYC 2892-1195-1 & -206.0 & 43:49:06.5 & \multicolumn{2}{c}{}  &  0&06     & 0&40 & \multicolumn{2}{c}{}  & \multicolumn{2}{c}{}  &   0.8 & 1.4          &            &       \\
    & GSC 02892-01195 &        &            & \multicolumn{2}{c}{}  &  0&35     & 0&45 & \multicolumn{2}{c}{}  & \multicolumn{2}{c}{}  & -37.0 & 3.0          &            &       \\
[1 ex]
 -- & 0431398+435319  &  830.6 & 04:31:39.8 & 11&35$_T$             &  0&37$_T$ & 0&47 & 10&76                 &  0&21                 &  -3.0 &  1.6         & B8 V *     & $pm$ \\
    & TYC 2892-1801-1 &  328.0 & 43:53:19.4 & \multicolumn{2}{c}{}  &  0&05$_C$ & 0&35 & \multicolumn{2}{c}{}  & \multicolumn{2}{c}{}  &  -0.1 &  1.6         &            &       \\
    & GSC 02892-01801 &        &            & \multicolumn{2}{c}{}  &  0&49$_C$ & 0&59 & \multicolumn{2}{c}{}  & \multicolumn{2}{c}{}  & -25.0 & 10.0         &            &       \\
[1 ex]
 -- & 0431025+434321  & 1684.0 & 04:31:02.5 & 11&43$_T$             &  0&37$_T$ & 0&32 & 10&89                 &  0&13                 &  -2.2 & 1.8          & A2 V       & $nm$  \\
    & TYC 2892-1171-1 & -967.2 & 43:43:21.2 & \multicolumn{2}{c}{}  &  0&28$_C$ & 0&23 & \multicolumn{2}{c}{}  & \multicolumn{2}{c}{}  &  -1.8 & 1.8          &            &       \\
    & --              &        &            & \multicolumn{2}{c}{}  &  0&45$_C$ & 0&39 & \multicolumn{2}{c}{}  & \multicolumn{2}{c}{}  & -21.0 & 5.0          &            &       \\
[1 ex]
  5 & J0432023+435228 &  306.1 & 04:32:02.3 & 11&53                 &  0&41                   & 0&35                   & 10&71                 &  0&22                 &  -1.7 & 2.0          & --         & $pm$ \\
    & TYC 2892-225-1  &  223.0 & 43:52:28.1 & \multicolumn{2}{c}{}  &  0&36                   & 0&26                   & \multicolumn{2}{c}{}  & \multicolumn{2}{c}{}  & -15.2 & 1.9          &            &       \\
    & --              &        &            & \multicolumn{2}{c}{}  &  0&49                   & \multicolumn{2}{c}{--} & \multicolumn{2}{c}{}  & \multicolumn{2}{c}{}  & \multicolumn{2}{l}{~~~~~~~--} &   &       \\
[1 ex]
  6 & J0432288+435210 & -309.4 & 04:32:28.8 & 11&93                 &  0&71                   & \multicolumn{2}{c}{--} & 10&62                 &  0&44                 & \multicolumn{2}{l}{~~~~~~~--} & -- & $nm$  \\
    & --              &  190.4 & 43:52:10.2 & \multicolumn{2}{c}{}  &  0&31                   & \multicolumn{2}{c}{--} & \multicolumn{2}{c}{}  & \multicolumn{2}{c}{}  & \multicolumn{2}{l}{~~~~~~~--} &    &       \\
    & --              &        &            & \multicolumn{2}{c}{}  &  0&80                   & \multicolumn{2}{c}{--} & \multicolumn{2}{c}{}  & \multicolumn{2}{c}{}  & \multicolumn{2}{l}{~~~~~~~--} &    &       \\
[1 ex]
  7 & J0432183+435149 &  -66.2 & 04:32:18.3 & 12&00                 &  0&45                   & 0&35                   & 11&11                 &  0&22                 & -5.8 & 1.9           & --         & $pm$ \\
    & TYC 2892-1159-1 &  143.8 & 43:51:49.6 & \multicolumn{2}{c}{}  &  0&35                   & 0&26                   & \multicolumn{2}{c}{}  & \multicolumn{2}{c}{}  &  1.0 & 1.8           &            &       \\
    & --              &        &            & \multicolumn{2}{c}{}  &  0&51                   & \multicolumn{2}{c}{--} & \multicolumn{2}{c}{}  & \multicolumn{2}{c}{}  & \multicolumn{2}{l}{~~~~~~~--} &   &       \\
\hline
\end{tabular}
\begin{minipage}{16cm}
\vspace{0.1cm}
{\bf Notes:}
\vspace{-0.3cm}
\begin{itemize}
\item Letters $T$, $H$ and $C$ indicate data obtained from Tycho-2, Hipparcos 
      catalog or computed according to the spectral classification (see 
      text) respectively.
\item Proper motion and radial velocity values are expressed in mas/yr and km/s
      respectively.
\item Spectral classification ($SC$) was obtained from Asiago observations.
      Asterisks indicate probable binary stars.
\item Membership (Memb.) is assigned as described in Sect.~3.4.
\end{itemize}      
\end{minipage}
\end{center}
\end{table*}

\begin{figure*}
\centering
\includegraphics[width=16cm]{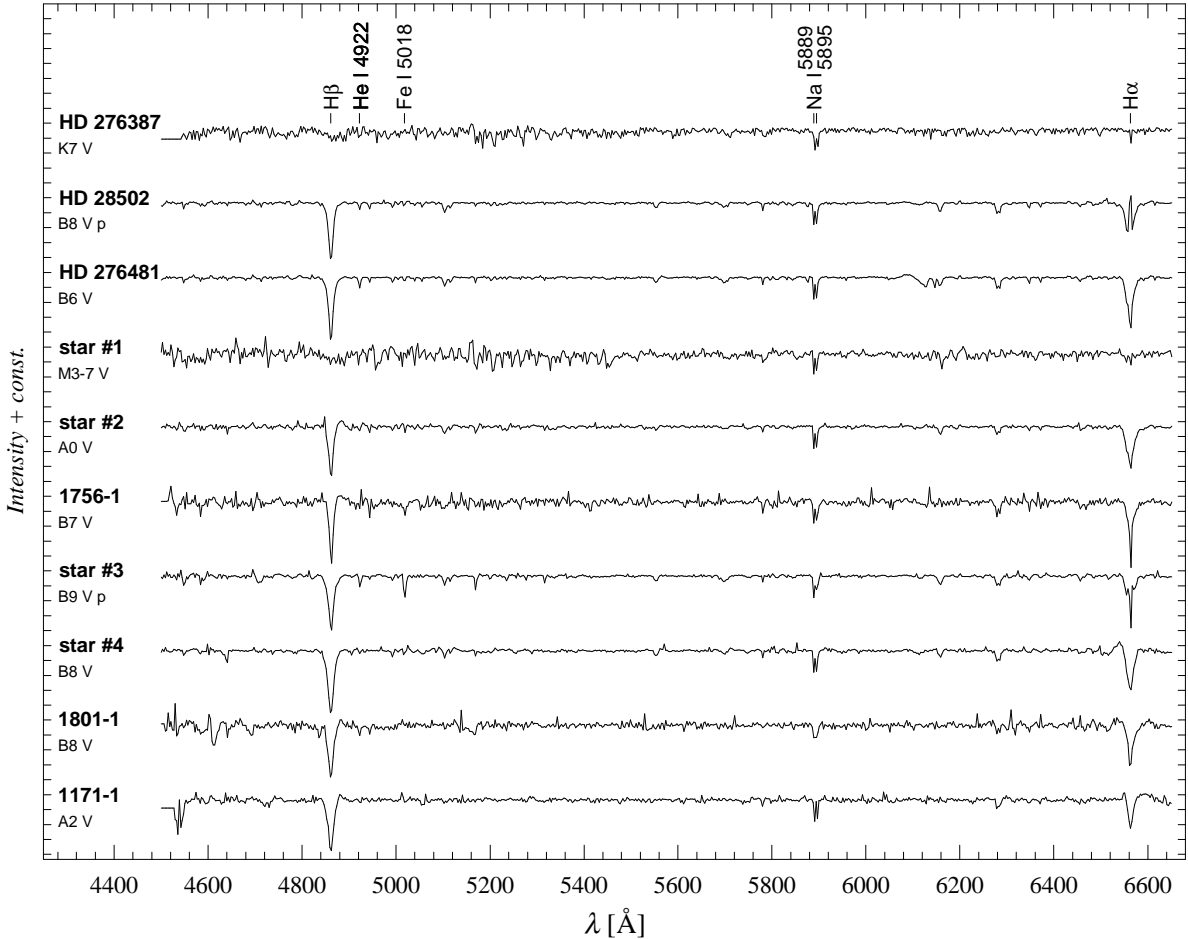}
\caption{{}Spectra of a sample of bright stars in the field of NGC~1582.
A few interesting lines are indicated. See Table~3 for details.}
\end{figure*}

\subsection{Proper motions and radial velocities}

$\hspace{0.5cm}$
Important information on the kinematics of the brightest stars in and around 
NGC 1582 field can be derived from the study of proper motions and radial 
velocities. The former are available in the Tycho-2 catalog and in the HD 
Extension Charts (Nesterov et al. 1995) whereas the latter were measured for 10 
stars (see Sect.~2.2). \\
\noindent
The Tycho-2 proper motions are based on the comparison between contemporary mean 
positions derived from the recent Tycho observations on-board Hipparcos and 
early-epoch positions observed many decades ago (see H{\o}g et al. 2000 and 
references therein). Due to the long time-baseline they have rather high 
precision and therefore directly indicate the long-term mean tangential motions 
of the stars. We have collected proper motion components for 53 stars in a field 
$20^{\prime}$ around the center of NGC~1582. They are shown in the vector point 
diagram in Fig.~4a. The points distribution is characterized by a global spread 
$\sim 10$~mas/yr with a noticeable concentration, which seems to indicate a 
possible physical relation among these stars. \\
\noindent
Spectra obtained with the Echelle spectrograph allow us to get radial velocity,
as described in  Sect.~2.2. The obtained values are shown in Fig.~4b, and
range from -40 to +40~km/s. \\

\begin{figure*}
\centering
\includegraphics[height=9cm]{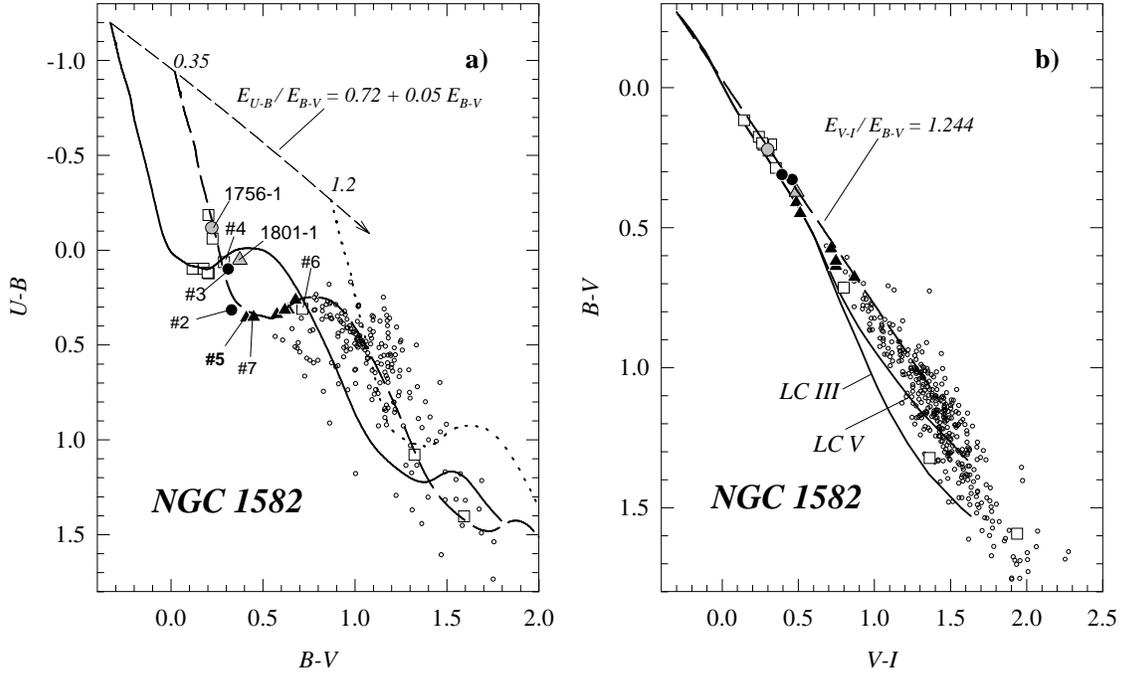}
\caption{{} Color-color diagrams (CCDs) of stars in the region of NGC~1582. 
{\bf a)} $U-B$ vs. $B-V$ diagram. The symbols have the following meaning: big 
circles are adopted member stars ($m$), triangles are probable member stars 
($pm$), empty squares are non-member stars ($nm$), and small open circles are 
stars without any membership assignment. Black symbols indicate stars with CCD
measurements whereas grey ones are for stars with data from the Tycho-2 catalog. The 
solid line is Schmidt-Kaler's (1982) ZAMS, whereas the dashed and dotted 
lines are the same ZAMS, but shifted by $E_{B-V} = 0.35$ and $1.2$, respectively. 
The dashed arrow indicates the normal reddening path. {\bf b)} $B-V$ vs. $V-I$ 
diagram. Symbols as in Fig. 5a. The solid lines represent the intrinsic 
positions for stars of luminosity classes V and III (Cousins 1978a,b). The dashed 
line gives the normal reddening path ($R = 3.1$).}
\end{figure*}

\subsection{Photometric diagrams}

$\hspace{0.5cm}$
The color-color diagrams (CCDs) and the color-magnitude diagrams (CMDs) are 
shown in Figs.~6, 7 and 8. The first two figures include all the stars measured 
in the direction of NGC~1582 and also several bright stars with available 
spectral classification and Hipparcos (ESA 1997) or Tycho-2 magnitudes whithin 
$20^{\prime}$ from the cluster center and not covered by our photometry. Tycho-2 
magnitudes are converted to the Johnson system using the relations given by
Bessell (2000) and approximated $(U-B)$ and $(V-I)$ colors are obtained 
according to the spectral types (when they are available) together with 
Schmidt-Kaler (1982) and Cousins (1978a,b) calibrations. Fig.~8 presents the CMDs 
from 2MASS catalog for stars placed inside the cluster radius ($R < 15^{\prime}$, 
see Sect.~3.1) and for star placed in a ring around the cluster ($20^{\prime} 
< R < 25^{\prime}$, see Fig.~3) that is adopted as a comparison field. Radii are 
selected in such a way that both diagrams in Fig.~8 cover equal sky areas. \\

\noindent
If we inspect Fig.~6a and compare the star positions with Schmidt-Kaler's (1982) 
Zero Age Main Sequence (ZAMS), using different reddening values, there seem to 
be two populations: one having a lower excess (dashed curve) mainly defined 
by the brightest stars, and another one with a much larger reddening. We claim 
that the former define the open cluster NGC~1582, whereas the latter population 
is identified as the Galactic disk component, made of stars placed at different 
distances and with a different amounts of absorption. To guide the eye we have 
placed another ZAMS reddened by $E_{B-V} = 1.2$ (dotted curve). A similar 
conclusion can be deduced from Figs.~7 and 8, where we see that most of the 
stars observed are just Galactic disk field stars and NGC~1582 looks like a 
small group of stars brighter than $V \approx 14$ above the mean stellar 
background. \\

\begin{figure*}
\centering
\includegraphics[height=9cm]{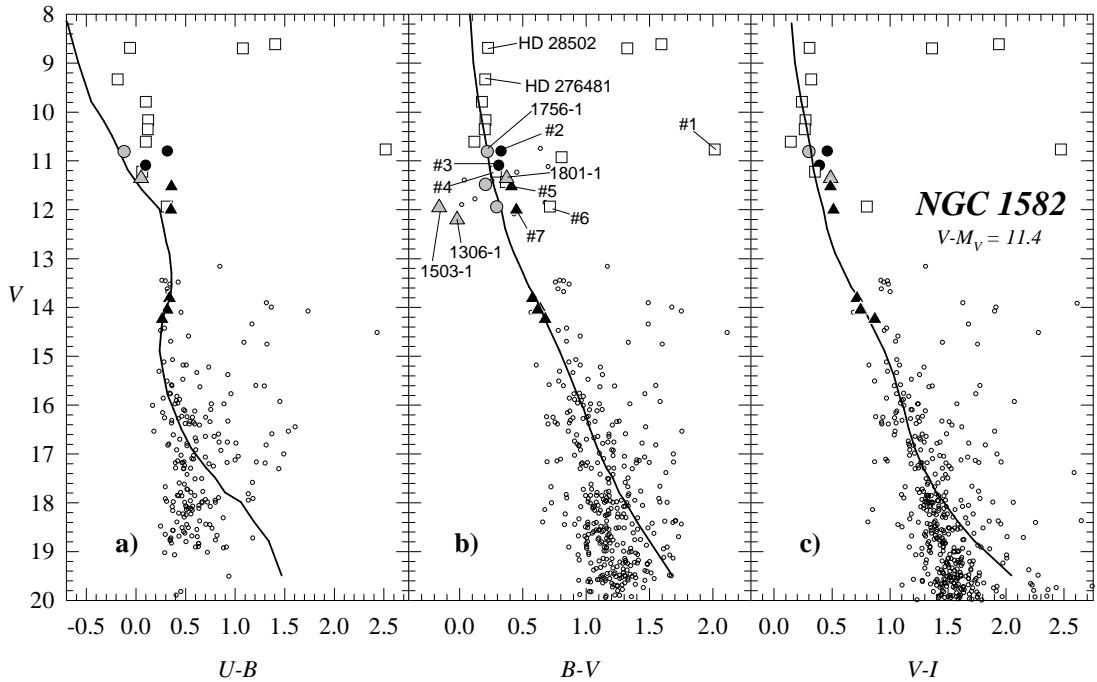}
\caption{{} CMDs for all the stars covered in the field of NGC 1582. Symbols as 
in Fig. 5a. The solid line is the Schmidt-Kaler (1982) empirical MS fitted to 
the apparent distance modulus $V-M_{V} = 11.4$ ($V-M_{V} = V_{0}-M_{V} + 3.1 
E_{B-V}$, see Sect.~3.4).}
\end{figure*}

\subsection{Members selection}

$\hspace{0.5cm}$
We derive cluster membership by comparing the distribution of the stars in
the different photometric diagrams (e.g. Baume et al. 1999, Ortolani et al. 
2002, Carraro 2002b). But we also take into account their location in the 
finding chart (Fig.~2), in the vector point diagram and the measured radial 
velocities (Fig.~4). At first, we use only Figs.~6 and 7 and we notice that 
there are five bright blue stars (\#2, 3, 4, 5 and 7) that fit pretty well the 
empirical ZAMS shifted by $E_{B-V} = 0.35$ (dashed line). These stars also are 
well placed in the vector point diagram, although three of them, stars \#4,  5
and 7, slightly depart from the central concentration in this diagram. 
Additionally, star \#4 has a radial velocity value very different from that of 
stars \#2 and 3. Therefore, stars \#2 and 3 are adopted as cluster members 
($m$), stars \#5 and 7 as probable members ($pm$), and star \#4 is considered to 
not belong to the cluster ($nm$). \\
Using stars \#2, 3, 5 and 7, we compute colour excess values for each of them
applying the relations $E_{U-B}/E_{B-V} = 0.72 + 0.05~E_{B-V}$ and $(U-B)_0 = 
3.69~(B-V)_0$ according to  V\'azquez \& Feinstein (1991), this yields a 
mean value $E_{B-V} = 0.35 \pm 0.03~(s.d.)$, here adopted as the cluster color 
excess. Beside, following the reddened ZAMS path onto CCDs and CMDs, we select 
four additional stars and adopt them as probable cluster members ($pm$). \\
\noindent
Star \#2 is of spectral type A0 V, as derived from its spectrum and from its 
position in the CCD. However by comparing its position in the CMDs with respect 
to other members, we notice that it appears over-luminous by $0.75$~mag. This 
can be explained by assuming it is a binary system. Of course, this is only
a suggestion, which can be confirmed for instance by looking for radial velocity
variations in other-epoch spectra.
Therefore we keep it on the adopted members list. On the other hand, star \#6 is 
well placed on the CCDs (see Fig~6) as a F-type star, but its position in the 
CMDs (see Fig.~7) contradicts this hypothesis and it is therefore considered a 
non-member ($nm$). \\
\noindent
At this point, we are ready to use Fig.~8 as well. In this figure we superimpose 
to the data an empirical Main Sequence (MS) obtained by the combination of the 
Schmidt-Kaler (1982) and Koornneef (1983) calibrations shifted according to the 
relation among colors obtained from the van de Hulst extinction curve \#15 
(Johnson 1968). By closely inspecting this figure, we notice that there are some 
stars not covered by our survey, but within the cluster radius, which are 
properly located both in the CMD (Fig.~8a) and in the proper motion diagram of 
Fig.~4a (grey symbols in those figures), and one of them (TYC 2892-1756-1) also 
has a radial velocity value compatible with those of adopted member stars \#2 
and 3 (see Fig.~4b). This group of stars has a counterpart in the corresponding 
comparison field (Fig.~8b). We compute then the $J$ distributions for stars with 
$J-K < 0.2$ (see Table~4) in each diagram and we compare them by using a $\chi^2$ 
test. We find that they are different with a probability higher than 95~\%. 
Therefore, although we are aware that we are dealing with  small numbers of 
stars and that a statistical analysis is therefore only indicative, the difference
between both distributions turns out to be noticeable and therefore we are 
inclined to consider the group of stars mentioned above as cluster members ($m$). An 
exception are stars TYC 2892-1306-1, TYC 2892-1503-1 and TYC 2892-1801-1. The 
first two have high errors in their $B-V$ measurements and are situated leftward 
of the MS in Fig.~7b, and the third has a radial velocity value with a huge 
error and is far from the adopted values for cluster members (see Fig.~4b and 
Table~3). We cannot exclude that this might be due to a binary effect. Thus we 
consider these three particular stars as only probable members ($pm$). On the 
other hand, the photometric diagrams also reveal that some stars with available 
spectral classification and Tycho-2 data do belong to the field stellar 
population; they are therefore taken as non-members ($nm$). \\

\begin{figure*}
\centering
\includegraphics[height=9cm]{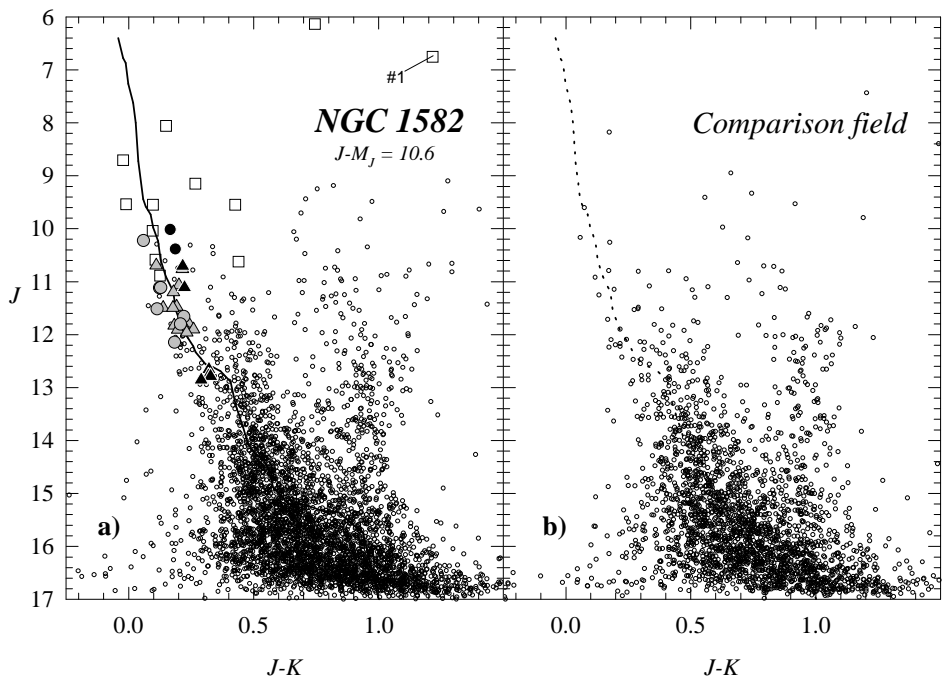}
\caption{{} CMDs from 2MASS catalog. Symbols are as in Fig.~6a. 
{\bf a)} Stars placed inside the cluster area. 
{\bf b)} Stars placed in a ring around the cluster. The solid line in panel {\bf a)}
and the dotted one in panel {\bf b)}
are the intrinsic position for MS stars from the Schmidt-Kaler (1982) and Koornneef 
(1983) calibrations fitted to the apparent distance modulus $J-M_{J} = 10.6$ 
($J-M_{J} = V_{0}-M_{V} + (3.1 - 2.3) E_{B-V}$, see Sect.~3.4).}
\end{figure*}

\noindent
Star \#1 (GSC 02892-01209) deserves special attention. It is the brightest 
and one of the reddest stars in our sample. Unfortunately, we were not able to 
find out either its distance or its proper motion components. This star might 
in principle be a giant cluster member. However, from its spectrum we obtained a 
radial velocity value that disagrees with the adopted one for adopted cluster 
members (see Fig~4b). Also, it is classified as a M-type star and its 
parameters (reddening and distance) differ from those of the cluster ones ruling 
out the possibility that it is a member star. \\

\noindent
In conclusion, after the above detailed analysis of the brightest stars in the
NGC~1582 region, we conclude that $i)$ the brightest ones are merely field 
stars, and $ii)$ we only consider the stars classified as $m$ and $pm$ as the 
main members of the open cluster NGC~1582. They are indicated in Fig.~2, and 
demonstrate that we are dealing with a sparse, poor and severely 
field-star-contaminated open cluster. \\

\subsection{Hints for distance and  age}

$\hspace{0.5cm}$
In Fig.~9 we plot the reddening-corrected $M_{V}$ vs. $(B-V)_{0}$ diagram 
for the cluster members and probable cluster members adopting a distance modulus 
of $V_{0}-M_{V} = 10.3 \pm 0.2$ (error from inspection). Last value fits very 
nicely the empirical Schmidt-Kaler (1982) ZAMS. We also apply the 
spectroscopic parallax method to three classified stars, obtaining 
$V_{0}-M_{V} = 9.9 \pm 0.2$. These values are in quite good agreement and
imply that the few stars identified as NGC~1582 are located $1100 \pm 100$~pc 
away from the Sun in the outer edge of the Orion arm. \\

From the obtained spectral classification (see Table~3) and from the location 
of adopted cluster members stars in Fig. 5a along the shifted ZAMS (dashed 
curve), we infer that the spectral types range from B7 to F2. If the stars 
having B7 spectral type are still on the main sequence, we derive an age of 
about $300$~Myr for NGC~1582 (Girardi et al. 2000). A similar result is obtained 
by the isochrone fitting method shown in Fig.~9. \\

\begin{table}
\caption{{} $J$ magnitude distributions from stars with $J-K < 0.2$ in Fig.~8.}
\begin{center}
\begin{tabular}{rccccc}
\hline
$\Delta J$         & 8-9 & 9-10 & 10-11 & 11-12 & 12-13 \\
\hline
$Cluster$          & 2   & 3    & 16    & 20    & 22    \\
$Comparison~field$ & 1   & 1    & 4     & 7     & 6     \\
\hline
\end{tabular}
\end{center}
\end{table}

\begin{figure}
\centering
\includegraphics[height=9cm]{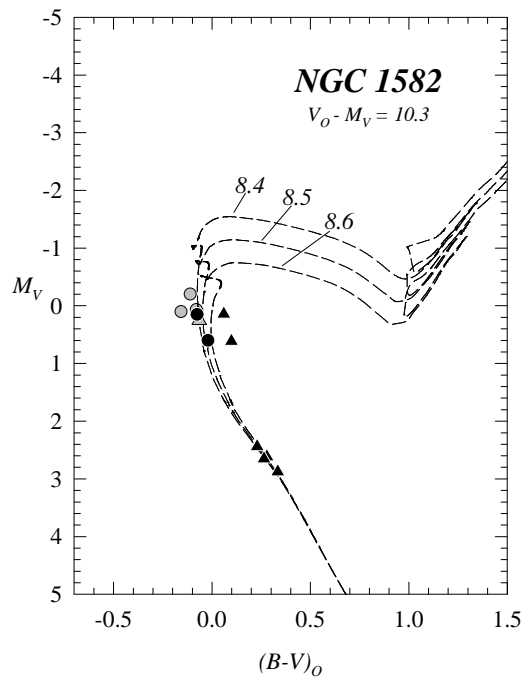}
\caption{{} $M_{V}$ vs. $(B-V)_{0}$ diagram of the members and probable members 
in the region of NGC~1582. Symbols as in Fig.~6a. Dashed curves are the 
isochrones from Girardi et al. (2000). The reported numbers give the 
$\log(age)$.}
\end{figure}

\section{NGC 1663}

\begin{figure}
\centering
\includegraphics[width=9cm]{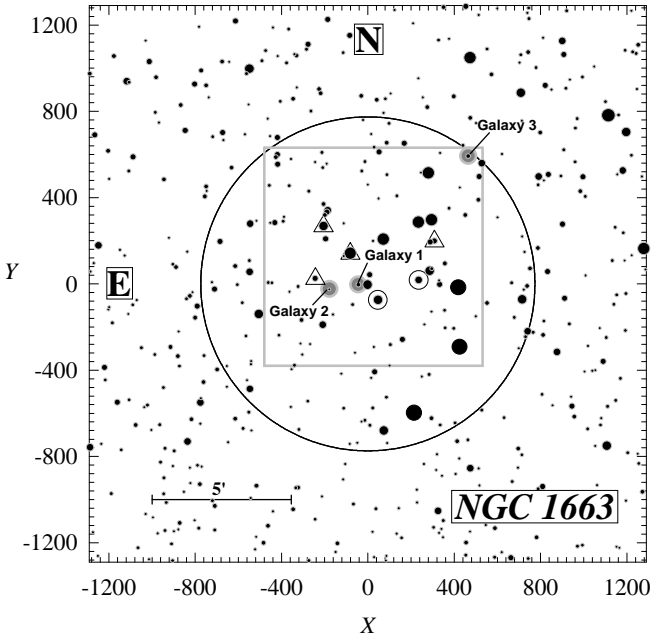}
\caption{{} Finding chart of the NGC~1663 region ($20^{\prime} \times 
20^{\prime}$ and $J$ filter). The black solid circle, $6^{\prime}$ radius, 
indicates the adopted angular size for the cluster (see Sect.~4.1 and Fig.~10). 
As in Fig.~2, grey square indicates the area covered from Asiago, and adopted 
members and probable members are enclosed in small circles and triangles, 
respectively. For a coordinate reference, the center ($X = 0$; $Y=0$) corresponds 
to the cluster coordinates (see Table~1), and each $X$-$Y$ unit is 
$0^{\prime\prime}.465$. Field galaxies positions are indicated as grey circles.}
\end{figure}

$\hspace{0.5cm}$
NGC~1663 (= OCL 461 = C0445+130) is located quite high above the galactic plane 
for an open cluster (see Table~1) and does not emerge much from the general 
Galactic field toward its direction. According to Lyng\aa\ (1987) it has a 
diameter of $12^{\prime}$ and therefore our observations cover most of the 
cluster region (see Fig.~10). Unfortunately, proper motions are available only 
for four stars in the region $10^{\prime}$ around NGC 1663 (see Table~5), 
so we must rely mostly on photometric data to derive cluster members and cluster 
fundamental parameters. \\

\subsection{Stellar counts}

$\hspace{0.5cm}$
NGC~1663 appears as a loose aggregate of a few relatively bright stars. As for 
NGC~1582, we perform stellar counts in concentric rings around the cluster 
center, but this time, since the cluster does not appear much extended, we only 
use the DSS-2 red data. The stellar density profile is shown in Fig.~11. One 
can readily see the lack of any clear trend, and therefore it is a difficult 
task to define the cluster radius. Anyway, if we inspect the DSS-2 image,  the 
region where is located the over-density of stars is about $6^{\prime}$ radius 
(see Fig.~10). In conclusion, it is hard to decide upon the real nature of this 
cluster, although, broadly speaking, the over-density we found seems to suggest 
that we are looking at an aggregate of the kind suggested by Bica et al. (2001), 
namely a POCR. The lack of an unambiguous cluster center, and the loose 
distribution of the brightest stars across the field support this suggestion. \\

\begin{figure}
\centering
\includegraphics[height=5.3cm]{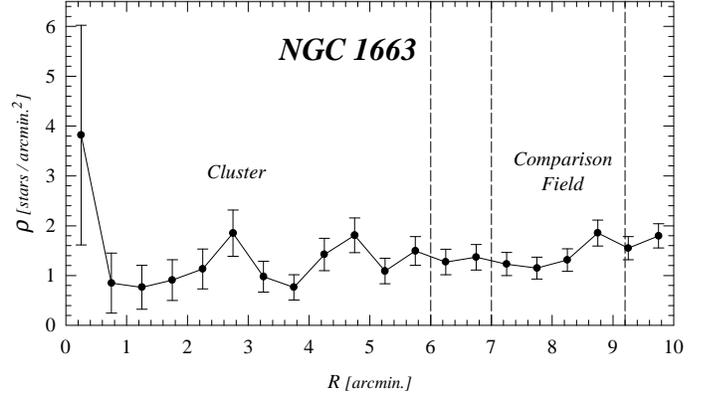}
\caption{{}Stellar densities in the region of NGC~1663 as a function of the 
radius from DSS-2 image. The dotted lines indicate the adopted limits for the 
cluster and for the comparison field related with Fig.~14b.} 
\end{figure}

\begin{figure*}
\centering
\includegraphics[height=9cm]{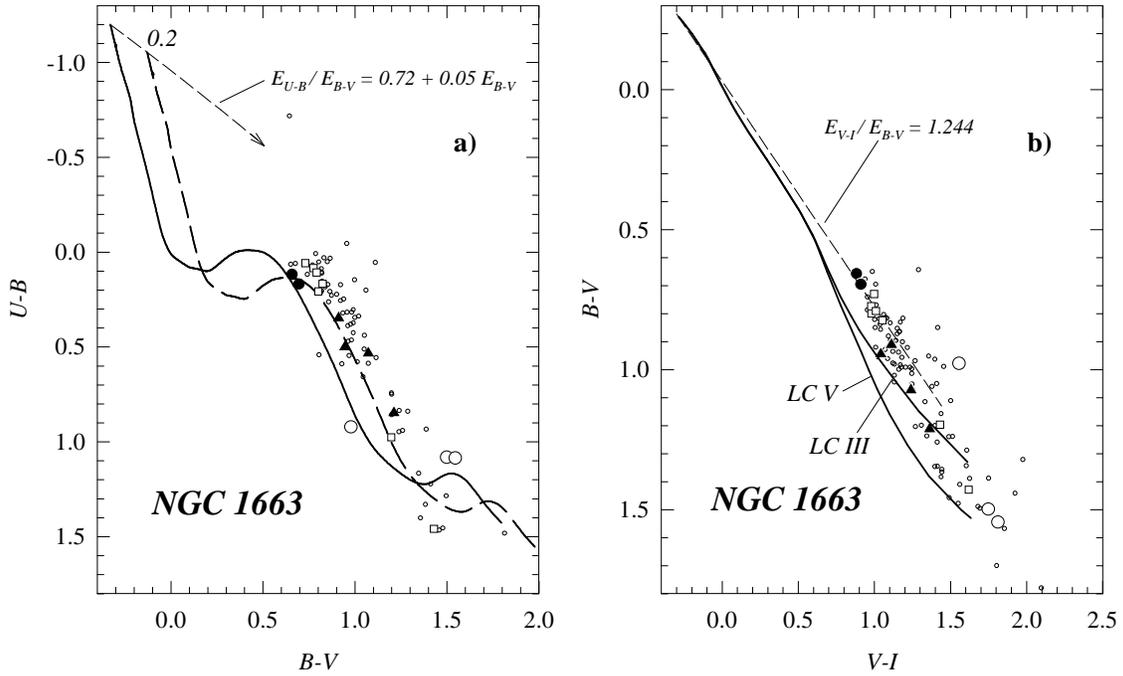}
\caption{{} CCDs of the stars in the region of NGC~1663. 
{\bf a)} $U-B$ vs. $B-V$ diagram. As for NGC~1582, symbols have the following 
meaning: circles are adopted member stars ($m$), triangles are probable member 
stars ($pm$), empty squares are non-members stars ($nm$) and small open 
circles are stars without any membership assignment. Field galaxies are 
indicated as white circles. The solid line is Schmidt-Kaler's (1982) ZAMS, 
whereas the dashed one indicates the same ZAMS, but shifted by $E_{B-V} = 0.2$. The 
dashed arrow indicates the normal reddening path. {\bf b)} $B-V$ vs. $V-I$ 
diagram. Symbols as in Fig.~12a. The solid lines are the intrinsic positions 
for stars of luminosity classes V and III (Cousins 1978a,b), while the dashed one 
provides the normal reddening path ($R = 3.1$).}
\end{figure*}

\begin{table*}
\caption{{} Some bright stars in the field of NGC 1663.}
\fontsize{8} {10pt}\selectfont
\begin{center}
\begin{tabular}{rlrcr@{.}lr@{.}lr@{.}lr@{.}lr@{$\pm$}ll}
\hline
 \# & 2MASS ID.       & \multicolumn{1}{c}{$X$} & $\alpha_{2000}$ & \multicolumn{2}{c}{$V$} & \multicolumn{2}{c}{$B-V$} & \multicolumn{2}{c}{$J$} & \multicolumn{2}{c}{$J-K$} & \multicolumn{2}{c} {$\mu_{\alpha} \cos(\delta)$} & Memb. \\
    & Tycho-2 ID.     & \multicolumn{1}{c}{$Y$} & $\delta_{2000}$ & \multicolumn{2}{c}{}    & \multicolumn{2}{c}{$U-B$} & \multicolumn{2}{c}{}    & \multicolumn{2}{c}{}      & \multicolumn{2}{c} {$\mu_{\delta}$}              &       \\
    & HD/GSC ID.      &                         &                 & \multicolumn{2}{c}{}    & \multicolumn{2}{c}{$V-I$} & \multicolumn{2}{c}{}    & \multicolumn{2}{c}{}      & \multicolumn{2}{c} {$Parallax$}                  &       \\
\hline
  1 & J0449107+130612 &    426.5 &    04:49:10.7 &    10&48             &  1&20                 &   8&09               &  0&71                &   13.1 &  1.5                   & $nm$   \\
    & TYC 691-633-1   &   -292.5 &    13:06:12.1 & \multicolumn{2}{c}{} &  0&98                 & \multicolumn{2}{c}{} & \multicolumn{2}{c}{} &  -12.4 &  1.5                   &        \\
    & HD 287125       &          &               & \multicolumn{2}{c}{} &  1&43                 & \multicolumn{2}{c}{} & \multicolumn{2}{c}{} &  108.4 & 46.3                   &        \\
[1 ex]
 -- & J0448489+131432 &   1115.0 &    04:48:48.9 &    10&47$_T$         &  0&49$_T$               &   9&14               &  0&38                &   22.2 & 1.4                  &  --  \\
    & TYC 695-519-1   &    780.6 &    13:14:32.2 & \multicolumn{2}{c}{} & \multicolumn{2}{c}{--}  & \multicolumn{2}{c}{} & \multicolumn{2}{c}{} &  -36.8 & 1.4                  &      \\
    & --              &          &               & \multicolumn{2}{c}{} & \multicolumn{2}{c}{--}  & \multicolumn{2}{c}{} & \multicolumn{2}{c}{} & \multicolumn{2}{l}{~~~~~~~--} &      \\
[1 ex]
 -- & J0449174+130349 &    213.6 &    04:49:17.4 &    10&68$_T$         &  1&63$_T$               &   7&96               &  0&92                &    6.6 & 2.3                  & --   \\
    & TYC 691-133-1   &   -597.8 &    13:03:49.6 & \multicolumn{2}{c}{} & \multicolumn{2}{c}{--}  & \multicolumn{2}{c}{} & \multicolumn{2}{c}{} &   -2.5 & 2.3                  &      \\
    & --              &          &               & \multicolumn{2}{c}{} & \multicolumn{2}{c}{--}  & \multicolumn{2}{c}{} & \multicolumn{2}{c}{} & \multicolumn{2}{l}{~~~~~~~--} &      \\
[1 ex]
  2 & J0449109+130819 &    419.0 &    04:49:10.9 &    11&69             &  1&43                 &   8&98               &  0&83                &    3.5 & 3.1                  & $nm$   \\
    & TYC 695-27-1    &    -17.6 &    13:08:19.9 & \multicolumn{2}{c}{} &  1&46                 & \multicolumn{2}{c}{} & \multicolumn{2}{c}{} &   -1.4 & 3.1                  &        \\
    & GSC 00695-00027 &          &               & \multicolumn{2}{c}{} &  1&62                 & \multicolumn{2}{c}{} & \multicolumn{2}{c}{} & \multicolumn{2}{l}{~~~~~~~--} &    \\
[1 ex]
  5 & J0449308+131030 &   -205.2 &    04:49:30.8 &    13&13             &  1&07                 &  11&06               &  0&65                 & \multicolumn{2}{l}{~~~~~~~--} & $pm$   \\
    & --              &    266.9 &    13:10:30.7 & \multicolumn{2}{c}{} &  0&53                 & \multicolumn{2}{c}{} & \multicolumn{2}{c}{}  & \multicolumn{2}{l}{~~~~~~~--} &        \\
    & --              &          &               & \multicolumn{2}{c}{} &  1&24                 & \multicolumn{2}{c}{} & \multicolumn{2}{c}{}  & \multicolumn{2}{l}{~~~~~~~--} &        \\
[1 ex]
  6 & J0449269+130932 &    -81.3 &    04:49:26.9 &    13&19             &  1&21                 &  10&88               &  0&69                 &  \multicolumn{2}{l}{~~~~~~~--} & $pm$   \\
    & --              &    140.3 &    13:09:32.1 & \multicolumn{2}{c}{} &  0&85                 & \multicolumn{2}{c}{} & \multicolumn{2}{c}{}  &  \multicolumn{2}{l}{~~~~~~~--} &        \\
    & --              &          &               & \multicolumn{2}{c}{} &  1&36                 & \multicolumn{2}{c}{} & \multicolumn{2}{c}{}  &  \multicolumn{2}{l}{~~~~~~~--} &        \\
[1 ex]
  7 & J0449227+130752 &     47.4 &    04:49:22.7 &    13&25             &  0&66                 &  11&79               &  0&39                 &  \multicolumn{2}{l}{~~~~~~~--} & $m$   \\
    & --              &    -74.5 &    13:07:52.2 & \multicolumn{2}{c}{} &  0&12                 & \multicolumn{2}{c}{} & \multicolumn{2}{c}{}  &  \multicolumn{2}{l}{~~~~~~~--} &        \\
    & --              &          &               & \multicolumn{2}{c}{} &  0&88                 & \multicolumn{2}{c}{} & \multicolumn{2}{c}{}  &  \multicolumn{2}{l}{~~~~~~~--} &        \\
[1 ex]
 11 & J0449167+130835 &    235.7 &    04:49:16.7 &    13&91             &  0&70    &  12&44               &  0&40                & \multicolumn{2}{l}{~~~~~~~--} & $m$   \\
    & --              &     17.1 &    13:08:35.4 & \multicolumn{2}{c}{} &  0&17    & \multicolumn{2}{c}{} & \multicolumn{2}{c}{} & \multicolumn{2}{l}{~~~~~~~--} &        \\
    & --              &          &               & \multicolumn{2}{c}{} &  0&91    & \multicolumn{2}{c}{} & \multicolumn{2}{c}{} & \multicolumn{2}{l}{~~~~~~~--} &        \\
[1 ex]
 16 & J0449320+130837 &   -242.9 &    04:49:32.0 &    14&73             &  0&94    &  13&01               &  0&46                & \multicolumn{2}{l}{~~~~~~~--} & $pm$  \\
    & --              &     24.1 &    13:08:37.6 & \multicolumn{2}{c}{} &  0&50    & \multicolumn{2}{c}{} & \multicolumn{2}{c}{} & \multicolumn{2}{l}{~~~~~~~--} &       \\
    & --              &          &               & \multicolumn{2}{c}{} &  1&04    & \multicolumn{2}{c}{} & \multicolumn{2}{c}{} & \multicolumn{2}{l}{~~~~~~~--} &       \\
[1 ex]
 18 & J0449144+130959 &    308.6 &    04:49:14.4 &    14&91             &  0&91    &  13&07               &  0&52                & \multicolumn{2}{l}{~~~~~~~--} & $pm$   \\
    & --              &    197.7 &    13:09:59.5 & \multicolumn{2}{c}{} &  0&35    & \multicolumn{2}{c}{} & \multicolumn{2}{c}{} & \multicolumn{2}{l}{~~~~~~~--} &        \\
    & --              &          &               & \multicolumn{2}{c}{} &  1&11    & \multicolumn{2}{c}{} & \multicolumn{2}{c}{} & \multicolumn{2}{l}{~~~~~~~--} &        \\
\hline
\end{tabular}
\begin{minipage}{16cm}
\vspace{0.1cm}
\hspace{1cm} {\bf Notes:} \\

\vspace{-0.3cm}
\hspace{1.3cm} - Letters $T$ indicate data obtained from Tycho-2 catalog. \\

\vspace{-0.3cm}
\hspace{1.3cm} - Proper motion and parallax values are expressed in mas/yr 
                 and mas respectively. \\
\end{minipage}
\end{center}
\end{table*}

\begin{figure*}
\centering
\includegraphics[height=9cm]{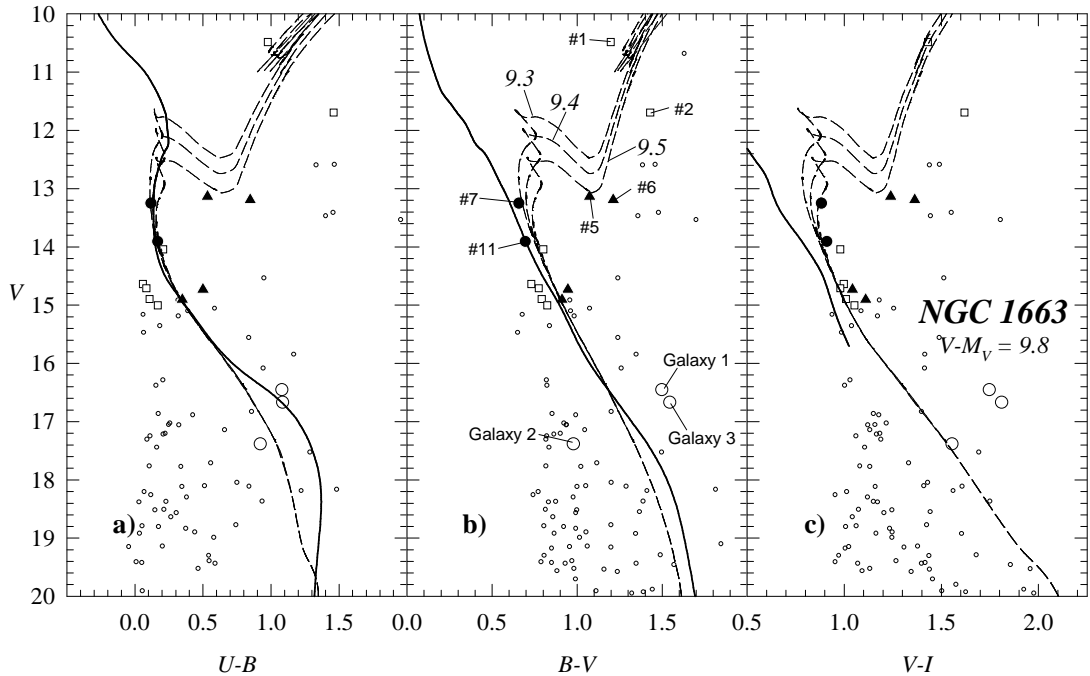} 
\caption{{} Color-magnitude diagrams (CMDs) for all the stars covered in the 
field of NGC 1663. Symbols as in Fig. 12a. The solid line and dashed curves are
the Schmidt-Kaler (1982) empirical ZAMS and isochrones from Girardi et al. 
(2002) respectively. They were fitted to the apparent distance modulus $V-M_{V} 
= 9.8$ ($V-M_{V} = V_{0}-M_{V} + 3.1 E_{B-V}$, see Sect.~4.2). The numbers 
indicate $\log(age)$.}
\end{figure*}

\begin{figure*}
\centering
\includegraphics[height=9cm]{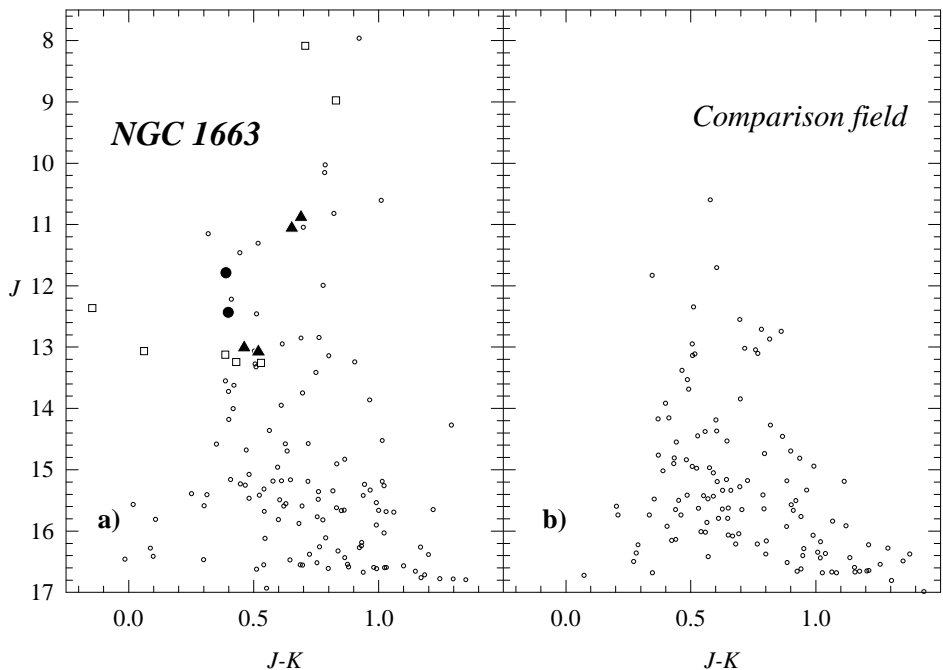}
\caption{{} CMDs from the 2MASS catalog. Symbols as in Fig. 12a. 
{\bf a)} Stars placed inside the cluster area {\bf b)} Stars placed in a ring 
around the cluster (see Fig. 10).}
\end{figure*}

\subsection{Photometric diagrams and member selection}

\begin{table*}
\caption{{} Galaxies observed in the field of NGC 1663}
\begin{center}
\begin{tabular}{lccccc}
\hline
{Name} & 
\multicolumn{1}{c} {$\alpha_{2000}$}  & 
$Morphology$ &
\multicolumn{1}{c} {$V$} & 
\multicolumn{1}{c} {$B-V$} & 
\multicolumn{1}{c} {$J$} \\
{2MASS ID} &
\multicolumn{1}{c} {$\delta_{2000}$} &
$D \times d$ &
&
\multicolumn{1}{c} {$U-B$}&
\multicolumn{1}{c} {$H$} \\
&&&&
\multicolumn{1}{c} {$V-I$}&
\multicolumn{1}{c} {$K$} \\
\hline

Galaxy 1      &  04:49:25.7 & $E3$                                           & 16.45 & 1.50 & 14.52 \\     
J0449257+1308 &  13:08:25.0 & $17.8^{\prime\prime}\times13.3^{\prime\prime}$ &       & 1.08 & 13.76 \\
              &             &                                                &       & 1.75 & 13.49 \\
[1 ex]   
Galaxy 2      &  04:49:29.9 & $S0$                                           & 17.38 & 0.98 & 16.33 \\
J0449299+1308 &  13:08:15.4 & $23.1^{\prime\prime}\times9.4^{\prime\prime}$  &       & 0.92 & 15.85 \\
              &             &                                                &       & 1.56 & 14.57 \\
[1 ex]
Galaxy 3      &  04:49:09.5 & $E0$                                           & 16.67 & 1.55 & 14.82 \\      
J0449095+1313 &  13:13:03.0 & $13.5^{\prime\prime}\times13.5^{\prime\prime}$ &       & 1.09 & 13.99 \\
              &             &                                                &       & 1.81 & 13.54 \\
\hline
\end{tabular}
\end{center}
\end{table*}

$\hspace{0.5cm}$
We follow the same method applied above for NGC~1582 to derive preliminary
individual reddening and membership of the cluster. Photometric diagrams are 
shown in Fig.~12, 13 and 14, where all the observed stars are presented in the 
first two, whereas the last one is obtained from 2MASS data. Let us fix our 
attention on Fig.~12a: here all the stars seem to crowd along an empirical ZAMS 
shifted by $E_{B-V} = 0.20$. It is however clear that in this case a reddening 
solution can not be found, since all the stars lie well beyond the location of 
A5 spectral type stars (Ortolani et al. 2002). The low color excess and the 
small dispersion in reddening are not unexpected, due to the position of the 
cluster high above the Galactic plane and toward the anti-center direction. The 
CMDs are not easy to interpret, since most of the stars are just Galactic disk 
field stars, as is readily seen by inspecting both panels in Fig.~14. From this 
figure it is evident that NGC~1663 emerges as an over-density of a dozen stars 
brighter than $V \approx 16-17$ above the mean stellar background. \\
\noindent
In conclusion, only preliminary membership information can be derived from the 
CMDs of Figs.~13 and 14. We therefore try to fit the distribution of the stars
with both the empirical Schmidt-Kaler (1982) ZAMS and with several isochrones 
from Girardi et al. (2000) by conservatively assuming that the metal content is 
solar. A possible relationship might exist between some MS stars and the 
brightest star (star \#1). This is the star HD~287125 with spectral 
type G5 and it could be a giant cluster member, but its parallax value (see 
Table~5) definitely rules out this possibility, and we are left with a nearby 
foreground star. The shape of the upper part of the CMDs seems to suggest that 
the turn-off point (TO) is located at about $V = 13.2$, $B-V = 0.6$, and that 
the stars rightwards of the TO are sub-giant stars. On this basis we will 
consider two stars as cluster members ($m$), four more as probable 
ones ($pm$) and a few bright ones were identified as non-members ($nm$). \\
\noindent
Fig.~13 shows the CMDs corresponding to the cluster area, and to a ring around 
it adopted as a comparison field (see Fig.~11). As in Fig~9 for NGC~1582, radii 
were selected in such a way that both diagrams in Fig.~14 represent equal sky 
areas. By comparing both diagrams we can see a notable over-density of bright 
stars in the cluster region, which would favor the idea that NGC~1663 is a 
physical cluster. \\

\subsection{Hints for NGC 1663 distance and age}

$\hspace{0.5cm}$
As discussed in the previous section, Fig.~13 allows us to derive a rough 
estimate of the distance and the age of NGC 1663. These diagrams yield a 
cluster distance of about $700$~pc ($V_{0}-M_{V} = 9.2 \pm 0.2$) and an age of 
about $2000$~Myr ($\log(age) = 9.4$). We also tried other combinations of 
distance moduli and color excesses over the CMDs, in order to check whether the 
reddest stars could be cluster members, but the obtained solutions did not 
agree at all with the corresponding CCDs of Fig.~12. \\
\noindent
In conclusion, if NGC 1663 really is a star cluster, we may be facing a 
dissolving aggregate of the kind proposed by Bica et al. (2001). \\

\section{Field galaxies}

$\hspace{0.5cm}$
Since NGC~1663 is located well above the Galactic plane, it is not very 
improbable that we should observe field galaxies toward its direction. In our 
case three of them are clearly detectable and we notice that they were not 
catalogued so far, except for entries in the 2MASS catalog as point sources. 
Therefore, we compute their integrated photometric parameters by performing 
aperture photometry. We use the PHOT task at increasing radius until the 
resulting magnitudes converged. Our results are presented in the photometric 
diagrams and in Table~6 together with an estimate of their angular sizes and a 
preliminary morphological classification that - by the way - agrees quite well 
with the observed colors. \\

\section{Conclusions}

$\hspace{0.5cm}$
We have presented the first CCD multicolor study in the regions of the two 
poorly known northern open clusters NGC~1582 and NGC~1663 for which no 
investigations had been carried out insofar. In the case of NGC~1582 we also 
obtained Echelle spectra of the brightest stars. In detail, we found 
that:
\begin{itemize}
\item NGC~1582 is a very poor and spread-out cluster with a radius of 
      15$^{\prime}$ and formed by a group of stars at a distance of about 1 kpc 
      in the outer edge of the Orion arm. We estimate that its reddening is
      $E_{B-V} = 0.35 \pm 0.03$ and that it has an age of about $300$~Myr. We 
      also obtained radial velocities for 10 stars and detected two probable 
      binary systems among its members.
\item NGC~1663 has a lower reddening of $E_{B-V} = 0.2$ but, until more robust 
      definition of its membership is available, the interpretation of the CMDs 
      is quite difficult. We only derive a preliminary membership assignment and 
      we suggest this object has an age of $\sim 2000$~Myr and a distance value 
      of about $700$~pc. The most probable interpretation of the data at our 
      disposal is that NGC~1663 is an open cluster remnant.
\end{itemize}

Finally, we identify three previously unclassified field galaxies in the 
direction of NGC~1663. We provide their integrated magnitudes and colors, 
angular sizes and preliminary morphological types. Two of them are found to be 
ellipticals and the third is either a flattened S0 or a spiral one. \\

{\it This article is partially based on the Second Generation Digitized Sky 
Survey that was produced at the Space Telescope Science Institute under US 
government grant NAG W-2166. The images of these surveys are based on 
photographic data obtained using the Oschin Telescope on Palomar Mountain and the 
UK Schmidt Telescope. The plates were processed into the present compressed 
digital form with the permission of these institutions. This study has also
made use of: a) the SIMBAD database, operated at CDS, Strasbourg, France, and
b) the data from the Two Micron All Sky Survey, which is a joint project
of the University of Massachusetts and the Infrared Processing and Analysis
Center, funded by NASA and NSF}

\begin{acknowledgements}
The authors acknowledge the Asiago Observatory staff for the technical 
assistance. Fruitful discussion with Roberto Barbon, Ruggero Stagni, Corrado 
Boeche and Silvano Desidera are also warmly acknowledged. We thank the 
anonymous referee for the detailed report which helped to significantly improve 
the paper presentation. The work of GB is supported by Padova University through
a postdoctoral grant.
\end{acknowledgements}

\end{document}